\documentclass[aps,twocolumn,showpacs,groupedaddress]{revtex4}
\usepackage{graphicx}

\begin{document}

\title{Experimentally realizable control fields in quantum Lyapunov control}
\author{X. X. Yi$^{1,2}$,  S. L. Wu$^1$, Chunfeng Wu$^2$, X. L. Feng$^2$, and C. H. Oh$^2$}
\affiliation{$^1$School of Physics and Optoelectronic Technology,\\
Dalian University of Technology, Dalian 116024 China\\
$^2$ Centre for Quantum Technologies and Department of Physics,
National University of Singapore, 117543, Singapore}

\date{\today}

\begin{abstract}
As a hybrid of techniques from open-loop and feedback control,
Lyapunov control has the advantage that it is free from the
measurement-induced decoherence but it  includes the system's
instantaneous message in the control loop. Often, the Lyapunov
control is confronted with time delay in the control fields and
difficulty in practical implementations of the control. In this
paper, we study the effect of time-delay on the Lyapunov control,
and explore the possibility of replacing the control field with a
pulse train or a bang-bang signal. The efficiency of the Lyapunov
control is also presented through examining the convergence time of
the controlled system. These results suggest that the Lyapunov
control is  robust gainst time delay, easy to realize and effective
for high-dimensional quantum systems.
\end{abstract}

 \maketitle
\section{introduction}
Quantum technologies, such as quantum information processing, offer
many advantages over their classical counterparts, but it is
challenging  to make these new technologies function robustly due to
the presence of noise. Quantum control---the application of control
theory to quantum systems---provides  a gateway  to develop robust
quantum technologies~\cite{wisemanmilburnbook,rabitz09}. Open-loop
control and closed-loop control (feedback control) are by definition
distinct but complementary control
methodologies~\cite{wisemanmilburnbook}. Open-loop control is
usually independent of measurement while feedback control, in
contrast, involves measurement. Both types of control have been
shown to be powerfully  experimental and theoretical tools in
classical and quantum contexts. For example,  optimal gate synthesis
by open-loop control~\cite{schulte05}, decoherence suppression by
bang-bang control and dynamical decoupling~\cite{viola99},
stabilization of pure and entangled states by feedback
control~\cite{wiseman05}, as well as the applications of feedback
control in precision metrology and hypothesis
testing~\cite{dolinar73}.

In spite of  some progress made, there are significant challenges
facing the field, in particular, the measurement-induced decoherence
in quantum feedback control  and that  stemming from
 experimental realizations of control fields in open-loop control.
Lyapunov control  is a hybrid of techniques from open-loop and
feedback controls, it  uses a feedback design to construct control
fields but  applies  the fields into the system in an open-loop way.
In other words, Lyapunov control is used to  design a feedback law
firstly which is then used to find the open-loop control by
simulating the closed-loop system; next the control is  applied to
the quantum system in an open-loop way. From the above description
of Lyapunov control, we find that the Lyapunov control includes two
steps. For any initial states and a system Hamiltonian (assumed to
be known exactly), the first step is to design a control law, i.e.,
to calculate the control field by simulating the dynamics of the
closed-loop system. The second step is to  apply the control law to
the control system as an open-loop control.

Although the Lyapunov control is not limited by the
measurement-induced decoherence, it is often confronted with
time-delays and realizability of the control fields in  practical
experiments. In this paper, we will shed light on these issues by
examining Lyapunov control on a finite dimensional system. The
following issues  will be clarified: (1) The robustness of Lyapunov
control against time-delay in the control fields; (2) the number of
pulses necessary to simulate the control fields; (3) the possibility
to replace the control fields with bang-bang signals, and (4) the
efficiency of the Lyapunov control.

The  paper is organized as follows. In Sec. {\rm II}, we introduce
the model and  explore the convergency of the Lyapunov control. In
Sec.{\rm III}, we examine the effect of time-delay on the control
fidelity. In Sec. {\rm IV}, we introduce a pulse train to simulate
the control field, and study the dependence of the fidelity on the
number of pulses. The possibility to replace the pulse train with
bang-bang signals and the efficiency of the Lyapunov control are
also explored in this section. Finally we end in Sec.{\rm V} with
discussions.

\section{Model description and convergency of the control}
Consider a quantum system  governed by ($\hbar=1$)
\begin{equation}
i\frac{\partial}{\partial t}|\psi(t)\rangle=(H_0+f(t)
H_1)|\psi(t)\rangle, \label{de}
\end{equation}
where $H_0$ and $H_1$ are $n\times n$ Hermitian matrices. Here $H_0$
is a time independent Hamiltonian, corresponding to  the free
evolution of the system in absence of any external fields. The
external interaction is taken into account as couplings $f(t)H_1$
between  a control field $f(t)$ and  the system (through a
time-independent Hermitian operator $H_1$). The state
$|\psi(t)\rangle$ is normalized,
$|\langle\psi(t)|\psi(t)\rangle|=1.$ Our goal is to steer an
arbitrary initial state $|\psi_0\rangle$ to the ground state of
$H_0$, say $|\phi_g\rangle.$ Although the ground state of $H_0$ is
specified as the target state in this paper, the present method is
not limited to this choice. In fact the method works for any target
states that are eigenstates of $H_0$. Define the following
real-value function $V(|\psi(t)\rangle, |\phi_g\rangle)$,
\begin{equation}
V(|\psi(t)\rangle,
|\phi_g\rangle)=1-|\langle\psi(t)|\phi_g\rangle|^2,
\end{equation}
we find $ \dot{V}=2f(t) \Im (\langle\phi_g|\psi(t)\rangle\langle
\psi(t)|H_1|\phi_g\rangle), $ where $\Im (...)$ denotes the
imaginary part of $(...).$ By choosing
\begin{equation}
f(t)=-k\Im (\langle\phi_g|\psi(t)\rangle\langle
\psi(t)|H_1|\phi_g\rangle),\label{cf1}
\end{equation}
with $k>0,$ $V$ decreases  along a trajectory that leads the system
from the initial state to the target state. Namely, any control
field of the form given in Eq. (\ref{cf1}) ensures $\dot{V}<0.$ With
such a control field $f(t)$, the distance between the actual state
$|\psi(t)\rangle$ and the goal state $|\phi_g\rangle$ decreases.

We prove the convergence of $|\psi(t)\rangle$ to the target  state
$|\phi_g\rangle$ (i.e., asymptotic stability) by showing  that the
LaSalle  invariant set $\mathcal{V}$ of states  satisfying  $\dot{V}
= 0$ does not contain trajectories of the system, except  the
trajectories that lead the system  to the target state. Clearly,
$\dot{V} = 0$ is equivalent to $\Im
(\langle\phi_g|\psi(t)\rangle\langle \psi(t)|H_1|\phi_g\rangle)=0.$
Notice that $\Im (ab) = 0$ if and only if there exists a real
numbers  $\lambda$ such that $\lambda a = b^*$. This is exactly the
case for $\Im (\langle\phi_g|\psi(t)\rangle\langle
\psi(t)|H_1|\phi_g\rangle)=0$, if $\langle \psi(t)| \phi_g\rangle
\neq 0.$ Therefore if $|\psi(t)\rangle$ belongs to $\mathcal V$,  it
must satisfy $\langle \psi_g|\lambda
\texttt{I}-H_1|\psi(t)\rangle=0$ or $\langle \psi(t)| \phi_g\rangle
=0,$ where $\texttt{I}$ denotes the identity operator. This means
that the  set  of states  orthogonal to $|\phi_g\rangle$ or being
eigenstates of $H_1$ must include the invariant set $\mathcal{V}$.
Namely, $\mathcal{V} \subseteq \{ |\psi(t)\rangle :\ \ \langle
\psi(t)| \phi_g\rangle =0 \ \ \text{or} \ \
H_1|\psi(t)\rangle=\lambda |\psi(t)\rangle \ \ \text{for any }
\lambda \}.$ For $\mathcal{V}$ to be {\it asymptotic} invariant, the
state in  $\mathcal{V}$ must be an eigenstate of $H_0,$ because the
free Hamiltonian can usually not be turned off. In other words, if a
state is not an eigenstate of $H_0$, it must evolve even if the
control field $f(t)$ is zero, and in consequence $f(t)$ would get a
non-zero value and thus the state would be outside the invariant
set. All these together yield that the LaSalle invariant set
$\mathcal{V} = \{ |\psi\rangle :\ \ H_0|\psi\rangle=E_0 |\psi\rangle
\ \ \text{for any } E_0\}.$ We
 call the states in the invariant set critical points of $V,$
which represent the maximum or minimum of the Lyapunov function $V.$
For a $n$-dimensional non-degenerate quantum system, there are  $n$
critical points, we now show that the maximal critical value occurs
only when $|\psi(t)\rangle=|\phi_g\rangle$.

Define operator  $A=\texttt{I}-|\phi_g\rangle\langle \phi_g|,$  the
Lyapunov function can be rewritten as $V(|\psi(t)\rangle,
|\phi_g\rangle)=\langle\psi(t)|A|\psi(t)\rangle.$ Obviously, $A$ and
$H_0$ have the same eigenstates. We denote the eigenstates of $A$ as
$|A_j\rangle$ ($j=1,2,...,n$) with corresponding eigenvalues $A_j.$
To determine the structure of $V(|\psi(t)\rangle, |\phi_g\rangle)$
around one of its critical points, say $|\psi_c\rangle=|A_c\rangle$
($c$ may be one of $1,2,...,n$), we consider an infinitesimal
variation $|\psi_c+\delta \psi\rangle$ of $| \psi_c\rangle$ such
that $\langle\psi_c+\delta \psi|\psi_c+\delta \psi\rangle=1.$
Express $|\psi_c+\delta \psi\rangle$ in the basis of the
eigenvectors of $A,$
\begin{eqnarray}
|\psi_c+\delta
\psi\rangle=|A_c\rangle+\sum_{j=1}^n\delta_{j}|A_{j}\rangle.\label{expan}
\end{eqnarray}
The normalization condition $\langle\psi_c+\delta \psi|\psi_c+\delta
\psi\rangle=1$ follows,
$$(\delta_c^{*}+\delta_c)+\sum_{j=1}^n\delta^{*}_{j}\delta_{j}=0.$$ Then
\begin{equation}
V(|\psi_c+\delta\psi\rangle)-V(|\psi_c\rangle)=\sum_{j\neq
c}(A_{j}-A_c)\delta_{j}^{*}\delta_{j}.
\end{equation}
Considering $\delta_{j}$ as variation parameters and noting
$\delta_{j}^{*}\delta_{j}\geq 0,$ we find that the structure of
$V(|\psi(t)\rangle, |\phi_g\rangle)$ around the critical point
$|\psi_c\rangle$ depends on the ordering of the eigenvalues of $A$.
$|\psi_c\rangle$ is a local maximum as a function of the variations
$\delta_{j}$ if and only if $A_c$ is the largest eigenvalue of $A$,
and a local minimum iff $A_c$ is the smallest eigenvalue and a
saddle point otherwise. This observation leads us to suspect that
the minimum of $V$ is asymptotically attractive, in other words, the
control field based on this Lyapunov function would drive the open
system to the eigenstate of $A$ with the smallest eigenvalue, which
is exactly the target state $|\phi_g\rangle.$

\section{time-delay effect}

For  practical implementation  of the Lyapunov-based control, a
non-negligible (positive or negative) time delay is required to be
taken into account. These delays  may stem from actuators and
electronic devices in the control loop \cite{nishio09}. Before
moving to the analytical discussions,   we first examine the effect
of time-delay on the fidelity of the Lyapunov control through an
example.

\begin{figure}
\includegraphics*[width=0.7\columnwidth,
height=0.5\columnwidth]{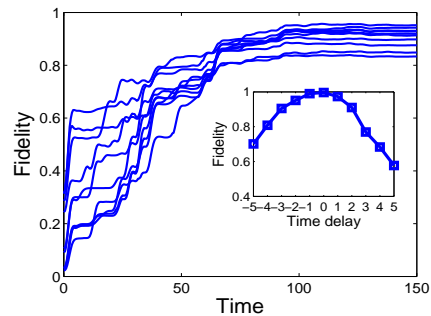} \caption{Fidelity as a
function of time for 10 different initial states with time delay
$\tau=1$. The initial states are randomly created by a MATLAB random
number generator, ignoring the relative phases between different
components. The inset shows the dependence of the average fidelity
on the time-delay. The average is taken over 10 randomly chosen
initial states. Throughout this paper, we rescale the maximal
eigenvalue of $H_0$ to be 1 and set $\hbar=1$, the time is rescaled
correspondingly. } \label{fig1}
\end{figure}

Take a 5-dimensional system as an example, and choose
\begin{equation}
H_0=\left(
\begin{array}{ccccc}
0.2&0&0&0&0\\
0&1.0&0&0&0\\
0&0&0.8&0&0\\
0&0&0&0.5&0\\
0&0&0&0&0.6\\
\end{array}
\right),\label{h0}
\end{equation}
as the free Hamiltonian. Let us use the previous Lyapunov control in
order to trap our system in the ground state
$|\phi_g\rangle=(1,0,0,0,0)^T$ of $H_0.$ Setting
\begin{equation}
H_1=\left(
\begin{array}{ccccc}
0&1&1&1&1\\
1&0&0&0&0\\
1&0&0&0&0\\
1&0&0&0&0\\
1&0&0&0&0\\
\end{array}
\right),\label{h1}
\end{equation}
as the control Hamiltonian and writing the actual state of the
system as
\begin{equation}
|\psi(t)\rangle=(\psi_1(t),\psi_2(t),\psi_3(t),\psi_4(t),\psi_5(t))^T,
\end{equation}
we obtain the control field by substituting $|\psi(t)\rangle,
|\phi_g\rangle,$ and $ H_1$ into Eq.(\ref{cf1}),
\begin{equation}
f(t)=-\Im[\psi_1(t)(\psi_2^*(t)+\psi_3^*(t)+\psi_4^*(t)+\psi_5^*(t))].\label{cfr}
\end{equation}
Consider Eq. (\ref{de}) without time-delay in $f(t)$, it has been
proved that \cite{mirrahimi04}, (1) if the spectrum of $H_0$ is not
degenerate, the invariant set of the closed loop system is spanned
by the eigenstates $|\Phi\rangle$ of $H_0$ such that
$\langle\Phi|H_1|\phi_g\rangle=0$; and (2)  when the invariant set
reduces to $\{|\phi_g\rangle, -|\phi_g\rangle \}$, the state
$|\phi_g\rangle$ is exponentially stable, while $-|\phi_g\rangle$ is
unstable. This means that the system would converge to
$|\phi_g\rangle$ under the control. Hence in the case of no
time-delay, the control field $f(t)$ perfectly steers the system
into the target state $|\phi_g\rangle.$ The situation changes when
the time delay is not zero. Note that the controlled dynamics
Eq.(\ref{de}) depends on the delay through the control field
$f(t-\tau)$ significantly, the control problem would be  much more
different than that without time-delay. We shall quantify the effect
of time delay on the controlled dynamics through
 control fidelity defined by,
\begin{equation}
F(t)=|\langle\phi_g|\psi(t)\rangle|^2.
\end{equation}
Fig.\ref{fig1} shows the control fidelity as a function of time and
the dependence  of final control fidelity on the time delay (see the
inset). We find that in the presence  of time delay, the final
control fidelity depends not only on the delay time, but also on
initial states. Moreover, we observe  that the negative time delay
and positive time delay are same for very small time delay but
different for large time delay, e.g., the control fidelity is
sensitive to positive time delays more than to the negative time
delay. This can be understood as follows. Consider a small time
delay $\tau$, the control field $f(t-\tau)$ can be expanded up to
second order in $\tau^2$ as
\begin{equation}
f(t-\tau)\simeq f(t)-\frac{\partial f(t)}{\partial t}\tau+\frac 1 2
\frac{\partial^2 f(t)}{\partial t^2} \tau^2.
\end{equation}
By simple algebra, we obtain,
$$ \frac{\partial f(t)}{\partial t}=ik\Im(\langle \phi_g|[H,|\psi(t)\rangle\langle \psi(t)|]H_1|\phi_g\rangle),$$
\begin{widetext}
$$ \frac{\partial^2 f(t)}{\partial t^2}=k\Im
(\langle\phi_g|H^2|\psi(t)\rangle\langle\psi(t)|H_1+|\psi(t)\rangle\langle
\psi(t)|H^2H_1-2H|\psi(t)\rangle\langle
\psi(t)|HH_1|\phi_g\rangle).$$
\end{widetext}
Without time delay, the system would approach  $|\phi_g\rangle$ as
time tends to the convergence time. This is not the case  when the
time delay is not zero. Suppose that the (un-normalized) final state
in the presence of time delay is
$|\psi(\infty)\rangle=|\phi_g\rangle+\frac{\lambda}{\sqrt{1+|\lambda|^2}}|\phi_g^{\bot}\rangle$,
where $|\phi_g^{\bot}\rangle$ denotes a state orthogonal to
$|\phi_g\rangle$ and $\lambda$ is a complex number.  It is easy to
show that $f(\infty)=-k\Im(\langle
\phi_g|\psi(\infty)\rangle\langle\psi(\infty)|H_1|\phi_g\rangle)=0,$
hence up to the first order in $\tau$, $|\phi_g^{\bot}\rangle$ does
not depend on $\tau$. This is the reason why the fidelity is
independent of $\tau$ when $\tau$ is very small. For large $\tau$,
e.g., the terms with $\tau^2$ are not negligible,
$|\phi_g^{\bot}\rangle$ can be calculated by $\frac{\partial
f}{\partial t}=\frac 1 2 \frac{\partial ^2 f}{\partial t^2}\tau$
with $|\psi(\infty)\rangle$ in  place of $|\psi(t)\rangle$. Clearly,
$|\phi_g^{\bot}\rangle$ depends on the delay time $\tau$, and
negative and positive time delays play different role.

\section{approaching the control field by a pulse train}
Observing Eq.(\ref{cf1}), we find that the control fields are a
continuous function of time. To realize such fields in experiment,
we need to manipulate the control fields at each point of time. This
is a challenging task in practical applications. The solution to
this problem is to replace the control field by a pulse train. Then
natural questions arise: How many pulses in the train are necessary
to simulate the control fields? Is the control fidelity sensitive to
the amplitude  and duration of the pulse? If not, can we replace the
Lyapunov control by a bang-bang control? In the following, we will
answer these questions.

\begin{figure}
\includegraphics*[width=0.7\columnwidth,
height=0.5\columnwidth]{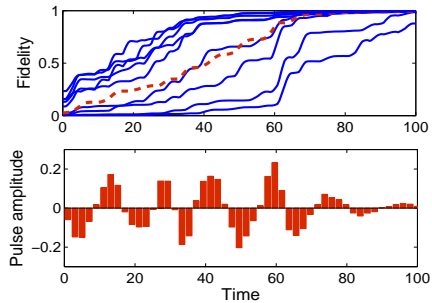} \caption{ Fidelity versus
time (upper panel), and pulse amplitude as a function of time (lower
panel). The initial states for the upper panel are randomly chosen,
while the control field shown in the lower panel corresponds to the
red-dot line in the upper panel. Note that 50 pulses are taken to
simulate  the control fields as shown in the lower panel.}
\label{fig2}
\end{figure}

\begin{figure}
\includegraphics*[width=0.7\columnwidth,
height=0.5\columnwidth]{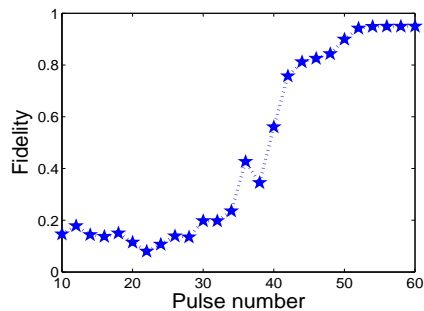} \caption{ Fidelity as a
function of pulse number. The fidelity is an averaged result taken
over 10 randomly chosen initial states.} \label{fig3}
\end{figure}

We numerically simulate the open-loop system Eqs.(\ref{de}),
(\ref{h0}), (\ref{h1}) and (\ref{cfr}) with a pulse train to replace
the control field $f(t)$. Numerical results are presented in
Fig.\ref{fig2}. Two observations can be made. (1) 50 sequence pulses
are enough for simulating the control fields, i.e., most initial
states can be driven  into the target state with 50 sequence pulses
in place of the control field. (2) The convergence time changes
(prolongs) when we use the pulse train to simulate the control
field. As expected, the fidelity depends on the number of pulse
significantly as Fig. \ref{fig3} shows, in particular around the
pulse number 40, the fidelity changes  abruptly. Of course, this
observation depends on the problem considered. The physics behind
these results can be understood as a wide range of $k$ allowed in
Eq. (\ref{cf1}). By properly choosing $k$ at each point of time, we
can re-arrange the control field and let it fit the pulse train.
When the duration of pulse in the train is long, it is difficult to
guarantee $k>0$, this degrades the fidelity of the control.

In control theory, a bang-bang control (on and off control) is a
feedback control that switches abruptly between two states.   In
optimal control problems, it is sometimes the case that a control is
restricted to be between a lower and an
 upper bound. If the optimal control switches from one extreme
 to the other at certain times (i.e., is never strictly in between the bounds)
then that control is referred to as a bang-bang one. The bang-band
controls  are  often implemented because of simplicity or
convenience. The physical idea behind bang-bang control comes from
refocusing techniques of NMR spectroscopy~\cite{ernst87}: control
cycles are implemented in time via a sequence of strong and rapid
pulses that provide a full decoupling from the environment. The
decoherence suppression results in the increase of the NMR
transversal relaxation time $T_2$, which is related to dephasing.
Besides NMR, dynamical decoupling has been suggested for inhibiting
the decay of unstable atomic states~\cite{kofman04,agarwal01},
suppressing the decoherence of magnetic states~\cite{search00}, and
reducing the heating in ion traps~\cite{vitali01}.

\begin{figure}
\includegraphics*[width=0.7\columnwidth,
height=0.5\columnwidth]{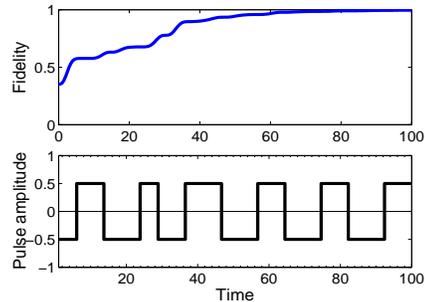} \caption{Pulse and fidelity
as a function of time. The initial state is randomly created, and in
this plot, it is $(0.4314,    0.3627,    0.5948,    0.3991,
0.4114)^T$. } \label{fig4}
\end{figure}

Noticing the simplicity and wide-range application of the bang-bang
control, one may wonder whether we can use the bang-bang signal to
replace  the control field $f(t)$. The answer is yes. Before going
to the detail, let us first examine Eq. (\ref{fig3}). The rate $k$
in the control field $f(t)$ implicates that the amplitude of the
control field is not important, in contrast, its sign plays an
exclusive role to guarantee $\dot{V}<0$. This is reminiscent of the
bang-bang signals and suggests us to arrange the control field in
the following way:
\begin{equation}
f(t)=\left \{
\begin{array}{cc}
f_0, & f(t)>0, \\
-f_0, & f(t)<0, \\
0, & f(t)=0,
\end{array}
\right.
\end{equation}
where $f_0$ is a constant. Fig. \ref{fig4} shows the fidelity as
well as the bang-bang signals as a function of time. We should
address that the duration of pulse and the time interval between two
pulses are different in the signal, which distinguishes this type of
bang-bang control from the traditional one.  Besides, the bang-bang
control here  differs from the traditional one at that  the design
of the control fields $f(t)$ is different. In the Lyapunov based
control, the bang-bang field is designed based on the Lyapunov
function, while in the traditional bang-bang control, the control
fields are designed based on the cost function.

\begin{figure}
\includegraphics*[width=0.7\columnwidth,
height=0.5\columnwidth]{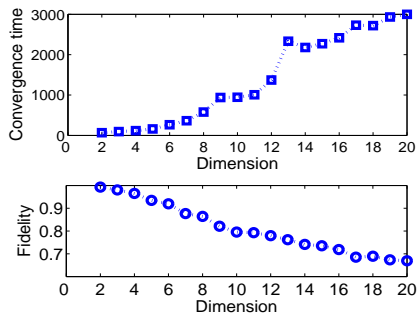} \caption{The convergence
time(upper panel) and the control fidelity (lower panel) versus
dimension of the control system. The convergence time is defined as
a time duration  with which the averaged fidelity arrives at or
larger than 0.95. The fidelity shown in the lower panel is taken at
time 150. All fidelities used here are averaged  over 100 randomly
created initial states.} \label{fig5}
\end{figure}

Before closing this section, we address the issue of efficiency of
the Lyapunov control. It is believed that the convergence time
increases as the dimension of the control system grows, but how does
the convergence time depend on the dimension of the system? The
convergence time here is defined as a time duration by which the
control system arrives at the target state with a fidelity of at
least $0.95$. Alternatively, the efficiency can also be described by
fidelities at which the system arrives within a fixed time duration.
We show both the convergence time and these fidelities as a function
of dimension in Fig.\ref{fig5}. The free Hamiltonian is randomly
created such that it is diagonal and its maximal element is set to
be 1, the control Hamiltonian is specified to be $H_1(1,1)=0,
H_1(m,1)=1 \ (m=2,3,4,...), H_1(1,n)=1 \ (n=2,3,4,...)$, and
$H_1(m,n)=0$ for the other $m$ and $n$. We find that the convergence
time prolongs almost linearly as the dimension of the system grows.
Accordingly, the fidelity degrades with the increase of the
dimension of the control system. This finding suggests  that
Lyapunov-based control becomes difficult for high dimensional
systems, but it is still an effective method to steer an open
quantum system.

\section{summary and discussions}
The Lyapunov  control fields are  traditionally required to
manipulate at each point of time. In this  paper, we have proposed a
scheme to replace  the Lyapunov control field by a pulse train or by
a bang-bang signal. Our analysis and numerical simulations show that
the control field can be well simulated by a pulse train, in which
each pulse has the same time duration but different amplitude. We
find that  the control fidelity depends on the number of pulses in
the train, and the dependence of the fidelity on the pulse number is
numerically given. The possibility to replace the Lyapunov control
field by a bang-bang signal is also explored. Surprisingly, the
bang-bang signal works well in the Lyapunov control to replace the
Lyapunov control fields. The physics  behind the replacement  is
given and analyzed.  The time-delay effect and the efficiency of the
control under the effect of the time-delay are also examined in this
paper. These results suggest that the Lyapunov control not only has
the advantage combining the feedback and open-loop control, but it
is also easy to realize for finite dimensional quantum systems.
\ \ \\
This work is supported by NSF of China under grant Nos 61078011 and
10935010, as well as the National Research Foundation and Ministry
of Education, Singapore under academic research grant No. WBS:
R-710-000-008-271.


\begin{references}

\bibitem{wisemanmilburnbook} H. M. Wiseman and G. J. Milburn, Quantum Measurement and
Control (Cambridge University Press, 2009).

\bibitem{rabitz09} H. Rabitz, New Journal of Physics 11, 105030 (2009).

\bibitem{schulte05} T. Schulte-Herbr\"uggen, A. Sp\"orl, N. Khaneja, and S. J.
Glaser, Phys. Rev. A 72, 042331 (2005); S. Sridharan, M. Gu, M. R.
James, ibid. 78, 052327 (2008).

\bibitem{viola99} L. Viola, E. Knill, and S. Lloyd, Phys. Rev. Lett. 82, 2417
(1999); G. Gordon, G. Kurizki and D. A. Lidar, ibid. 101, 010403
(2008)

\bibitem{wiseman05} H. M. Wiseman and A. C. Doherty, Phys. Rev. Lett. 94, 070405
(2005); M. Mirrahimi and R. Van Handel, SIAM J. Control Optim. 46,
445 (2007); A. R. R. Carvalho, J. J. Hope, Phys. Rev. A 76,
010301(R) (2007).

\bibitem{nishio09} K. Nishio, K. Kashima, and J. Imura, Phys. Rev. A {\bf 79}, 062105 (2009).

\bibitem{dolinar73} S. Dolinar, Tech. Rep. 111, Research Laboratory of Electronics,
MIT Technical Report No. 111, (1973); H. M. Wiseman and R. B.
Killip, Phys. Rev. A 57, 2169 (1998).


\bibitem{ernst87} R. Ernst, G. Bodenhausen, and A. Wokaun, Principles of Nuclear
Magnetic Resonance in One and Two Dimen- sions (Clarendon Press,
Oxford, 1987).

\bibitem{kofman04} A. G. Kofman and G. Kurizki, Phys. Rev. Lett. 93, 130406 (2004).

\bibitem{agarwal01} G. S. Agarwal, M. O. Scully, and H. Walther, Phys. Rev. Lett. 86,
4271 (2001).

\bibitem{search00} C. Search and P. R. Berman, Phys. Rev. Lett. 85, 2272 (2000).

\bibitem{vitali01} D. Vitali and P. Tombesi, Phys. Rev. A 65, 012305 (2001).

\bibitem{mirrahimi04} M. Mirrahimi, P. Rouchon, In proceedings of
IFAC symposium, NOLCOS's 2004.


\end{references}
\end{document}